\documentclass{pazheng}
\usepackage{amsmath}
\usepackage{amsfonts}
\usepackage{amssymb}
\usepackage{amstext}
\usepackage{natbib}
\usepackage{indentfirst}
\usepackage{misccorr}
\bibpunct{(}{)}{;}{a}{,}{,}
\usepackage{lscape}
\usepackage{mathtext}
\usepackage{graphicx}
\usepackage{float}
\usepackage{textcomp}
\usepackage{mathtext}
\usepackage{epsfig}
\usepackage{rotating}
\usepackage{dcolumn}
\usepackage{tabularx}
\newcommand{\dd}{$^\circ~$}
\newcommand{\ddm}{^\circ~}

\newcommand{\pih}{\frac{\pi}{2}}
\begin{document}
\title{Identification of X-ray lines in the spectrum of the arcsec-scale precessing jets of SS 433}
\author{\bf 
I.~Khabibullin\email{khabibullin@iki.rssi.ru}\address{1,2} and 
S.~Sazonov \address{1}
\addresstext{1}{Space Research Institute, Russian Academy of Sciences, Moscow}
\addresstext{2}{Max-Planck Institute for Astrophysics, Garching, Germany}  
}
\shortauthor{}
\shorttitle{}
\submitted{November 7, 2016}
\begin{abstract}

~~~~~The extended X-ray emission observed at arcsec scales along the
  propagation trajectory of the precessing relativistic jets of the
  Galactic microquasar SS 433 features a broad emission line, with the
  position of the centroid being significantly different for the
  approaching and receding jets ($ \approx7.3 $ and $ \approx6.4 $
  keV, respectively). These observed line positions are at odds with
  the predictions of the kinematic model for any of the plausible 
  bright spectral lines in this band, raising the question of their
  identification. Here we address this issue by taking into account
  time delays of the emission coming from the receding regions of the
  jets relative to that from the approaching ones, which cause a
  substantial phase shift and distortion of the predicted line
  positions for the extended ($ \sim 10^{17}$ cm) emission compared to
  the X-ray and optical lines observed from the central source
  (emitted at distances $ \sim 10^{11}$ cm and $ \sim 10^{15}$ cm, 
  respectively). We demonstrate that the observed line positions are
  fully consistent with the Fe XXVI Ly$\alpha$ ($E_0=6.96$  keV) line
  emerging from a region of size $ \sim6\times10^{16}$ cm along the
  jet. This supports the idea that intensive reheating of the jets up
  to temperatures $ \gtrsim10 $ keV takes place at these
  distances, probably as a result of partial deceleration of the jets
  due to interaction with the surrounding medium, which might cause 
  collisions between discrete dense blobs inside the jets.

\englishkeywords{black holes, accretion, jets, SS 433}
\end{abstract}    
%
\section{Introduction}
\label{s:intro}

~~~The most unique property of the Galactic microquasar SS 433 is
continuous launching of the highly collimated (with the opening angle
$ \Theta_j\approx1.5$\dd), mildly relativistic (with the bulk velocity
$ V=\beta c\approx0.26c $), baryonic (i.e. composed of atoms of hydrogen,
helium and heavy elements with the relative proportion close to the
solar one) jets, which reveal themselves via Doppler-shifted optical
and X-ray emission lines detected in the spectrum of the unresolved
core of the system (see \cite{Fabrika2004} for a review). The Doppler
shifts of the lines periodically vary with time in accord with the
kinematic model \citep{Abell1979,Fabian1979,Milgrom1979} postulating
that the jets undergo precessional motion about an axis inclined at $
i=78.05\pm0.05\approx$78\dd relative to the line of sight with
half-amplitude $\Theta_p=20.92\pm0.08\approx $21\dd and period
$P=162.375\pm0.011\approx162$ days (\citealt{Eikenberry2001}).

~~Apart from the unresolved core, the presence of the jets is clearly
traced by radio emission at distances from $ \sim 10$ milliarcseconds
(e.g. \cite{Marshall2013}) to a few arcseconds, where it forms a
remarkable 'corkscrew' pattern generally consistent with the
prediction of the kinematic model (e.g. \citep{Miller2008}), albeit
with slight deflections that might be caused by the interaction of
the jets with the surrounding medium \citep{Panferov2014}. Further
away, however, the jets stay invisible up to $ \sim 10 $ arcminute
scales, where extended X-ray emission, which probably has relation to
their propagation through the interstellar medium (ISM), is observed
\citep{Brinkmann2005}, and then they appear to be terminated at the
boundary of the surrounding radio nebula W50 \citep{Dubner1998},
giving rise to its severe deformation \citep{Goodall2011} and bright
X-ray emission from the supposed termination region
\citep{Brinkmann1996}.   
 
~~~Thanks to the excellent angular resolution of the \textit{Chandra}
X-ray observatory ($\approx 0.5$ arcsec), \cite{Marshall2002} have
discovered extended X-ray emission at a distance of the order of one
arcsec from the central source, coming from regions coincident with
the jets' propagation trajectory as traced by the radio emission 
observed at the same angular scales. Moreover, in further
observations, oppositely Doppler-shifted emission lines were detected 
in the spectra of the (on average) approaching (hereafter the 'blue'
jet) and receding (hereafter the 'red' jet) jets, with the lines'
centroids at $\approx 7.3 $ keV and $ \approx 6.4 $ keV, respectively
\citep{Migliari2002}, which indicates that the lines originate in the
material of the jets, still moving at a mildly relativistic speed.

~~~However, the exact identification of these lines remains elusive:
using simulations of the line emission of the extended jets averaged
over a precession period, \cite{Migliari2002} argued that the
rest-frame energy of the lines could be 7.06 keV. In such a case, however,
the line should be identified with the Fe I $  K\beta $ line of
neutral iron rather than with the Fe XXV $K\beta $ line (7.89 keV) of
highly ionized iron, as suggested by \cite{Migliari2002}. While the
presence of the Fe XXV line is a clear indication of hot ($ \gtrsim$
few keV) plasma, the Fe I line arises as a result of fluorescence
following inner shell ionization of neutral iron atoms by X-ray
photons or by interaction with energetic particles. The ambiguity in
the line identification leaves open the question about the mechanism
of production of the extended X-ray emission at arcsec scales,
corresponding to distances of order $\sim 10^{17} $ cm from the
central source. 

~~~In this paper, we address the identification issue of the extended
emission lines taking into account time delays between signals coming from
different parts of the precessing jets due to finite light propagation
speed, which prove to be important in predicting the line positions
for radiation emitted at distances of order $ \sim\lambda=\beta
cP=1.1\times 10^{17}$ cm from the central source, in contrast to the
well-known X-ray and optical lines emitted at $\sim10^{11} $ cm and
$\sim10^{15} $ cm, respectively (see \cite{Fabrika2004}). Indeed, at
distances $ \sim\lambda $, the time delay between different regions of
the jets can reach $ \sim2\lambda\tan\Theta_p/c\sim 32$ days, which
corresponds to a significant precession phase shift of $\approx
0.2$ for the observed emission. This effect is also important and has
previously been taken into account for modelling images and
variability of the radio emission observed at the same angular scales
(see e.g. \cite{Bell2011} and references therein).

~~~As a simplest scenario, we may assume that the arcsec X-ray line
emission is produced at some specific distance $ r $ from the central
source or, equivalently, at some specific instant after launch of the
emitting blob from the jet's core. As $r$ gets larger, the observed
line positions will increasingly deviate from the ones predicted by
the kinematic model without accounting for light-travel-time
effects, eventually leading to ambiguity in the predicted line positions
due to simultaneous arrival of emission from different parts of the jet.  
This will further lead to broadening and brightening of the observed
emission line if the emission is unresolved spatially and spectrally. 

~~~In a more complicated scenario, the X-ray lines are radiated over a
range of distances (or delays after launch) from the central source,
and the observed line positions are determined by properly averaging 
over the whole emitting region. In this case, they will still manifest
some modulation with the precession phase even for a brightening
duration comparable to the period of precession, as a result of
Doppler boosting of the emission from the approaching parts of the
jets relative to the receding ones. 

~~~We demonstrate that the observed line positions are consistent
either with the Fe XXV K$\alpha$ ($E_0=6.7$  keV) line of helium-like
iron in the former scenario, or with the Fe XXVI Ly$\alpha$ 
($E_0=6.96$  keV) line of hydrogen-like iron in the latter, hence with
a hot gas origin of the lines in both cases. However, the required
precession phase of the central source in the case of identification
with Fe XXV K$\alpha$ disagrees with the observed one, whereas there
is good agreement in the case of the Fe XXVI Ly$\alpha$ line.  

~~~Because quite high ($ \gtrsim 10$ keV) temperature is needed for
the Fe XXVI Ly$\alpha$ line to dominate over the Fe XXV K$\alpha$ line
in the thermal spectrum of hot plasma, the inferred identification
implies that the jets' material experiences efficient heating at
distances $ \sim 1.5\lambda $ from the central source lasting a
significant fraction ($ \sim 0.6$) of the precession period. This may
be caused by partial dissipation of the kinetic energy of the jets via
the propagation of a shock in their matter as a result of their
interaction with the surrounding medium (see,
e.g. \cite{Heavens1990}), or by mutual collisions of discrete dense 
blobs within the jets due to more effecient deceleration of the
leading blobs by the interstellar medium with respect to the trailing
ones.

~~~The paper is organised as follows: in the first section, we
describe a model for the variation of the line positions as a function of
the precession phase depending on location and size of their emission
region along the jet. In the second section, we compare predictions of
the model with the observed positions of the emission lines for the
approaching and receding jets and draw a conclusion regarding
their possible identification. The last section is devoted to the
discussion of the results obtained and final conclusions of the
present study. 

\section{Predicted line positions}
\label{s:model}
%
\begin{figure}[t]
\centering
\includegraphics[width=1.0\columnwidth,bb=50 280 550 680]{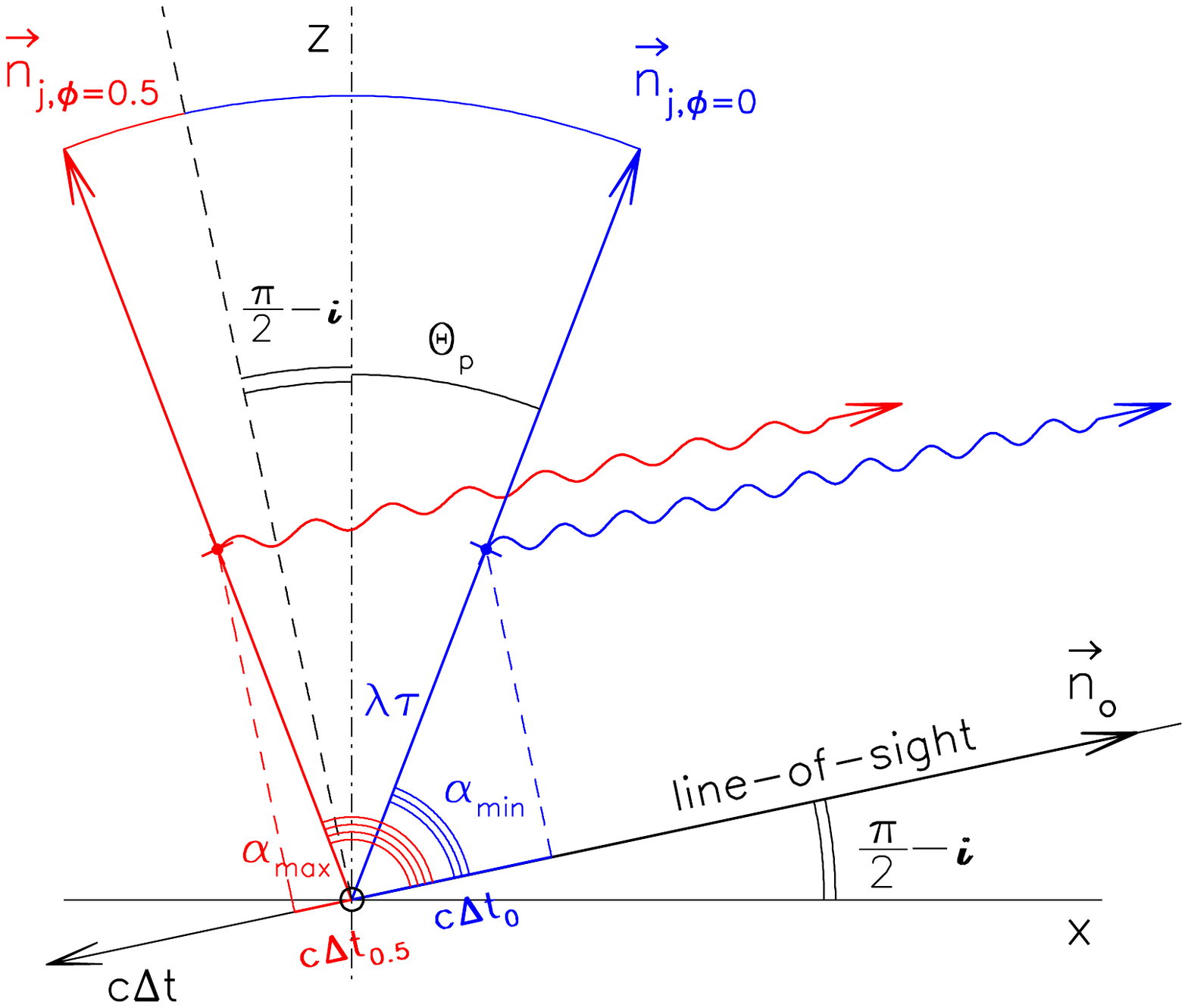}
\caption{\small Geometry of the problem for the approaching (on
  average) jet. The precession axis coincides with the $z$ axis, and
  together with the direction to the observer $\vec{n}_{o}$, inclined
  at angle $ i\approx78\ddm $ to the former, they form the $xz$
  plane. The direction of the jet $\vec{n}_{j}$ makes an angle $
  \Theta_p\approx21\ddm $ with the precession axis, and an angle
  ranging from $ \alpha_{min}=i-\Theta_p\approx58\ddm$ at precession
  phase $\phi=0$ (shown in blue) and $
  \alpha_{max}=i+\Theta_p\approx99\ddm$ at $\phi=0.5$ (shown in red) 
  with the line of sight. The black dashed line marks the projection
  of the picture plane. The differences in the distance to the
  observer from the approaching and receding parts of the jet relative to
  the distance from the central source are denoted by $c\Delta t_{0}$
  and $c\Delta t_{0.5}$ for the precession phases $\phi=0$ and
  $\phi=0.5$, respectively, and the regions of the 'blue' jet launched
  at $\phi=0.5$ are in fact receding from the observer.} 
\label{f:sketch}
\end{figure}

Let us first consider the innermost regions of the precessing
jets, whose observational appearance is not affected by
light-travel-time effects. It is sufficient to consider just one of 
the jets, e.g. the blue one, since the results for the other one can
be readily obtained by inversion of the jet's direction vector.   

Let the $z$ axis be aligned with the precession axis and the $x$ axis
lie in the plane formed by the $z$ axis and the direction from the
source to the observer. Then, the $y$ axis and the line in the $xz$
plane normal to the line of sight form the picture plane (see
Fig. \ref{f:sketch}). Hence, the line-of-sight and precession axis
directions are given by vectors 
\begin{equation}
\vec{n}_{o}=(\sin i,~0,~\cos i),~ \vec{n}_{p}=(0,~0,~1), 
\end{equation}
where $i$ is the inclination angle of the precession axis relative to the line of sight. 

The direction of the blue jet varies periodically in time and depends
on the precession phase as 
\begin{equation}
 \vec{n}_{j}(\phi)=(\sin\Theta_p\cos 2\pi\phi,~\sin\Theta_p\sin 2\pi\phi,~\cos\Theta_p),
\end{equation} 
where $ \Theta_p $ is the amplitude of precession, i.e. the
half-opening angle of the precession cone, $ P $ is the precession
period, and the zero precession phase is defined here as the moment
corresponding to the minimum angle $ \alpha_{min}=i-\Theta_p $ between
the jet's direction and the line of sight\footnote{Hereafter, the
  angle between the jet's direction and the line of sight is measured
  in the rest frame of the observer, so it does not need to be
  corrected for relativistic aberration, as would be the case for the
  angle measured in the reference frame comoving with the jet's
  matter. Time is also measured in the observer's reference frame
  and is thus not affected by relativistic dilation.}. The maximum
angle $ \alpha_{max}=i+\Theta_p $ then corresponds to $\phi=0.5$, and
since $\Theta_p>\pih-i$, the blue jet is actually receding at this
moment (see Fig. \ref{f:sketch}). 
\begin{equation}
\mu(\phi)=\cos i~\cos\Theta_p+\sin i~\sin\Theta_p\cos 2\pi\phi,
\end{equation}
reaching the maximum $\mu_{max}=\cos(i-\Theta_p)$ at $ \phi=0$ and minimum $ \mu_{min}=\cos(i+\Theta_p)$ at $ \phi=0.5$.

The Doppler factor for the jet's matter moving ballistically with the bulk velocity $ V=\beta c$ is given by 
\begin{equation}
\delta(\mu)=\frac{1}{\gamma(1-\mu(\phi)\beta)},
\end{equation}
with $ \gamma=1/\sqrt{1-\beta^2}$. Photons with energy $ E_0 $ emitted by the jet will be detected by the observer with the energy    
\begin{equation}
E(\phi)=\frac{E_0}{\gamma(1-\mu(\phi)\beta)},
\label{eq:e}
\end{equation}
and this equation indeed provides an excellent fit to the observed line positions for the compact (unresolved) jets in SS 433 (with the line positions for the red jet obtained by substituting $ \mu $ by $ -\mu $).

Suppose now that the jet's matter becomes bright (e.g. after having been reheated) at some instant $\tau P$ (expressed in units of the precession period $P$ through the dimensionless time-delay parameter $\tau$) after its launch from the central object. The corresponding distance covered by the jet by this time is $ \beta c~\tau P=\lambda\tau $, where $ \lambda=\beta c P$ is the distance covered by the jets over one precession period. Photons from the reheating region of the jets will reach the observer with a delay
\begin{equation}
\Delta t(\tau,\mu)=(1-\mu\beta)\tau P
\end{equation}
relative to photons emitted by the central source, where the $-\mu\beta\tau P$ component is caused by the finite light travel time.

Thus, if the observer cannot spatially separate the emission of the
central source from the radiation produced at distance $ \lambda\tau $
from it, the detected signal will contain both photons coming from the
central source at its current precession phase $ \phi_0 $ and from
blobs launched from the central source at an earlier phase $
\phi_e=\phi_0-\Delta\phi$, where     
\begin{equation}
\Delta\phi(\tau,\mu_e)=(1-\mu_e\beta)\tau, 
\end{equation}
and $\mu_e(\tau,\phi_0)$ (cosine of the angle between the line of
sight and the velocity direction of the matter emitting at
distance $ \lambda\tau $) is determined implicitly by the non-linear
equation 
\begin{equation}
\mu_e=\mu(\phi_0-\Delta\phi(\tau,\mu_e)).
\label{eq:mu}
\end{equation}
 
\begin{figure}
\centering
\includegraphics[width=0.4\textwidth,bb=80 200 530 680]{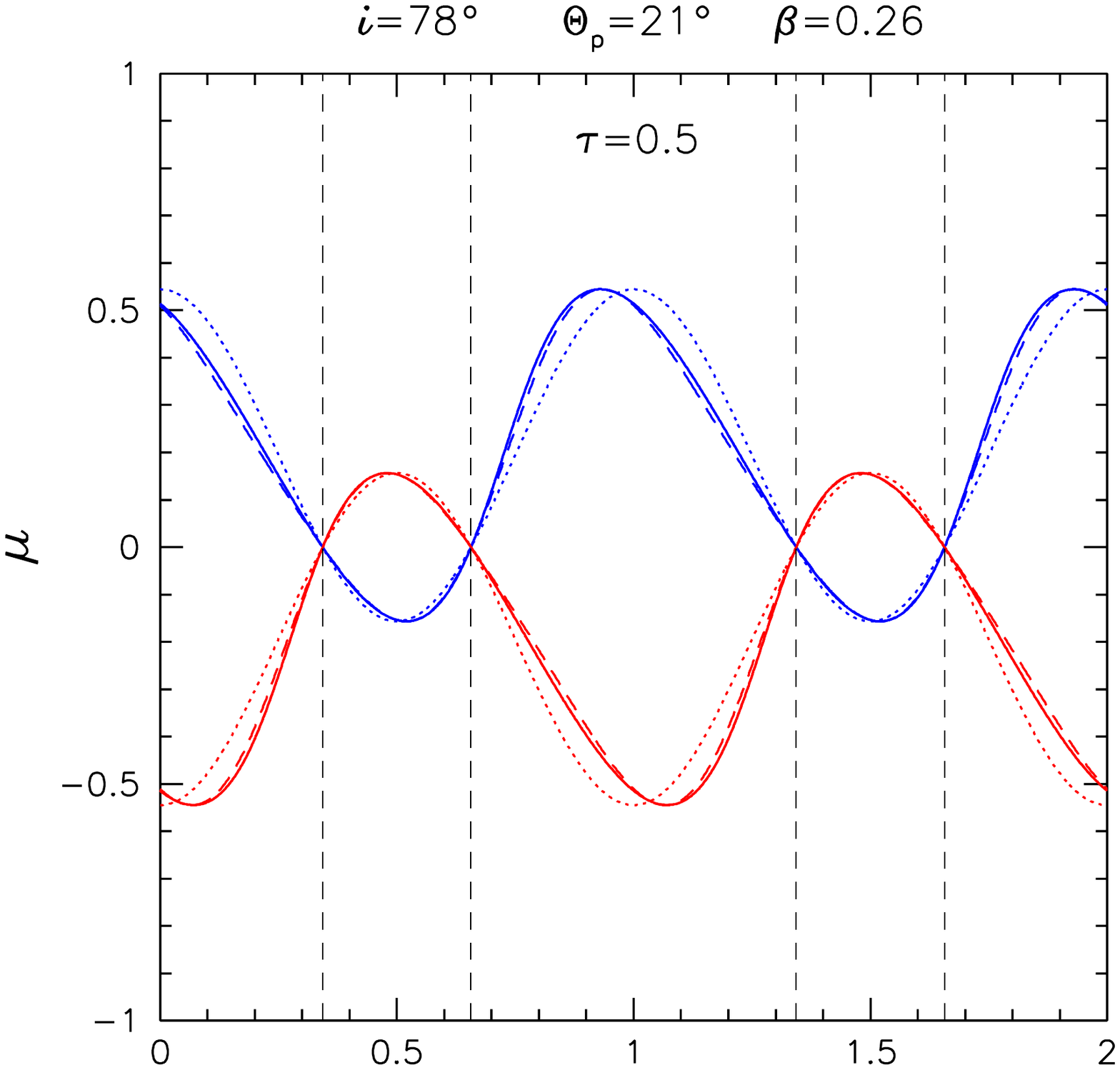}
\includegraphics[width=0.4\textwidth,bb=80 200 530 650]{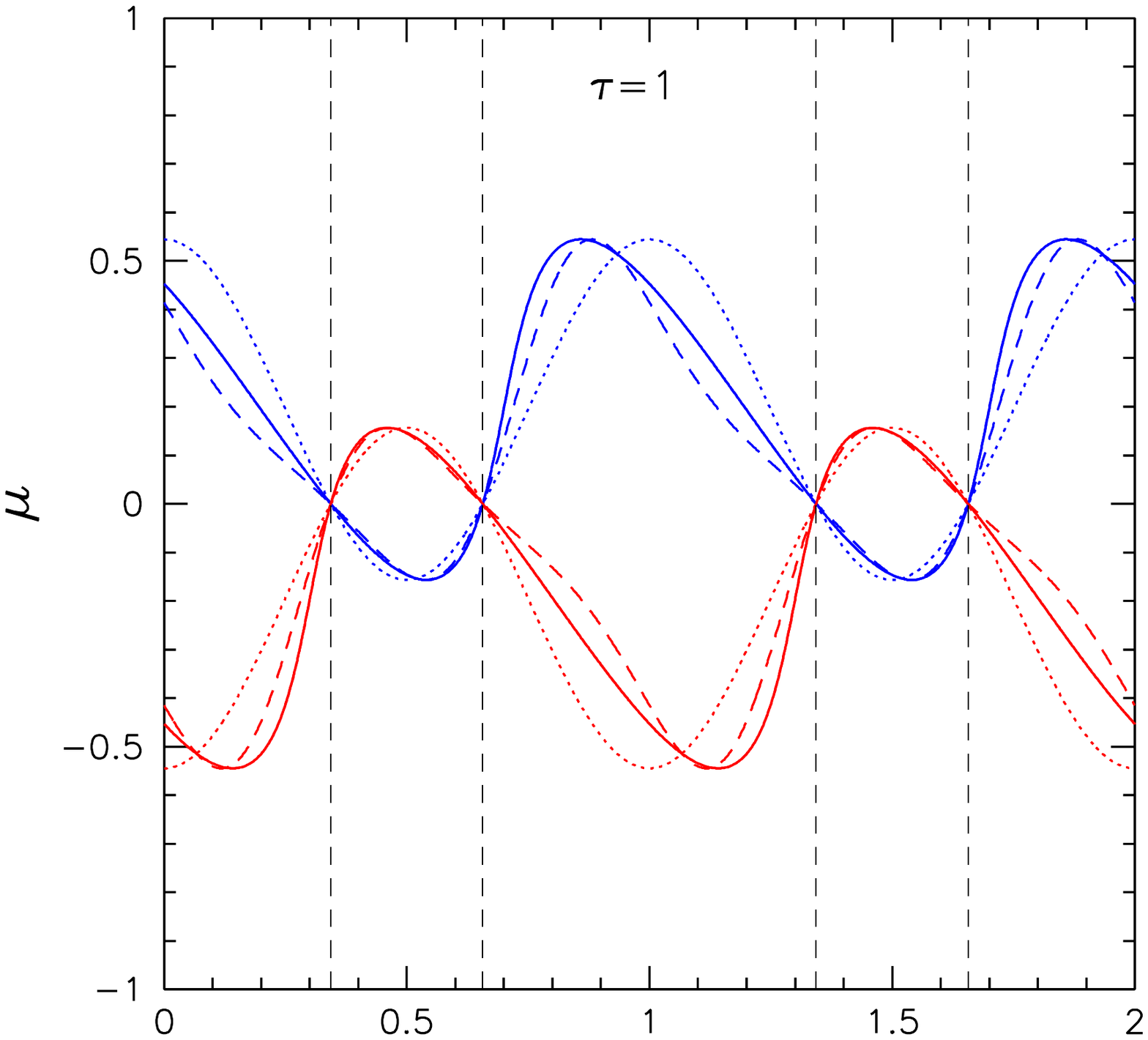}
\includegraphics[width=0.4\textwidth,bb=80 180 530 650]{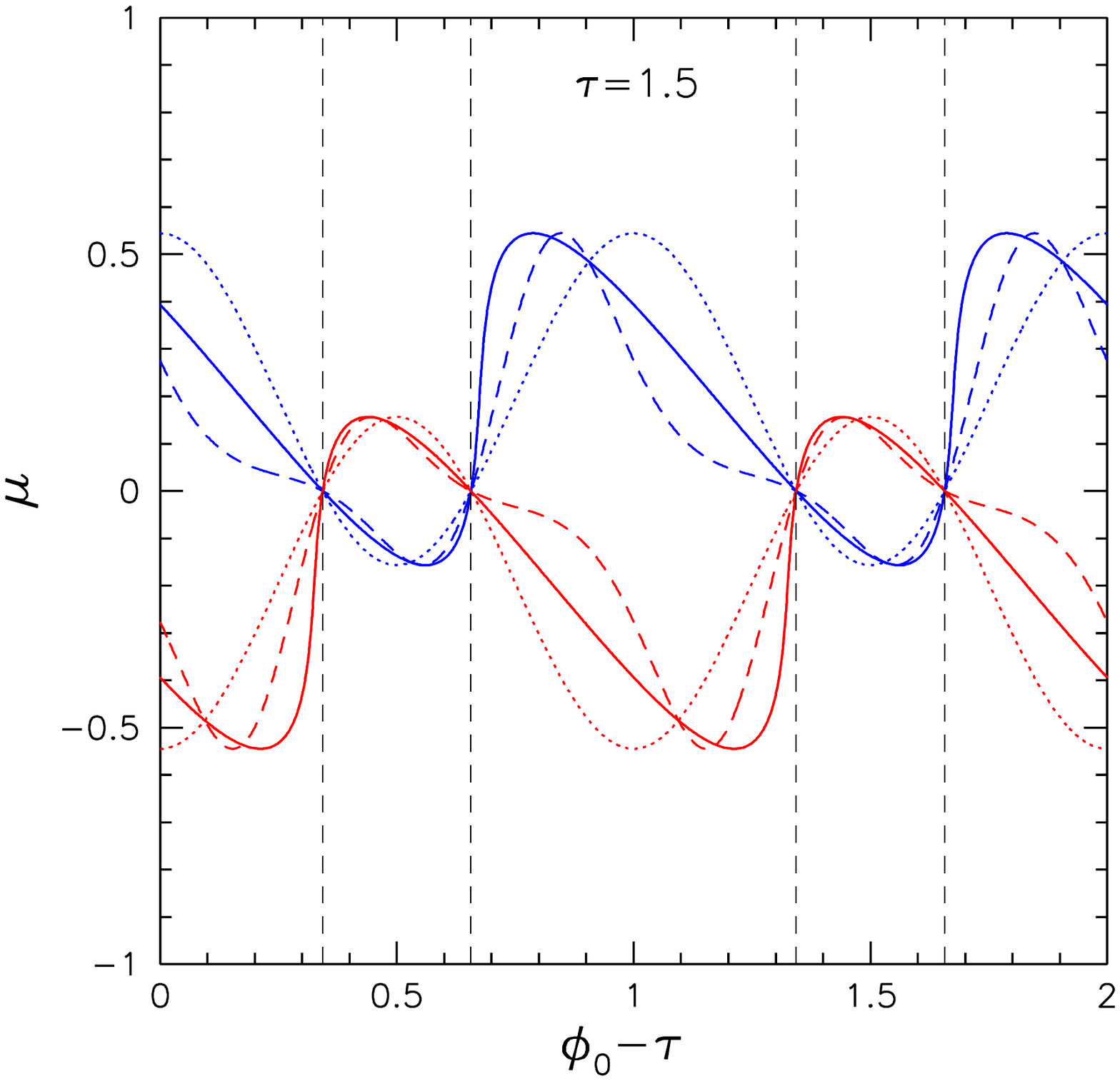}
\caption{\small Cosine of the angle between the observed jet's direction and the line of sight as a function of the current precession phase of the central source $ \phi_0 $ shifted by $ \tau $ for $ \tau=0.5,1.0,1.5 $ (from top to bottom). Model predictions for the blue and red jets are shown with blue and red solid lines, respectively. Variation of the line position for the compact jets is shown by the dotted line, while the dashed line shows the 'linear' approximation of the exact solution for the extended jets. The vertical dashed lines mark the stationary points corresponding to the jet's propagation in the picture plane.}
\label{f:mu}
\end{figure}
 
We see that the total phase shift $ \Delta\phi(\tau,\mu_e) $ is the sum of two components. It is the $\mu_e\tau\beta$ term that makes Eq. \ref{eq:mu} implicit, while the $\tau$ component just shifts the observed precession phase $ \phi_0 $ backward. When $ \beta\tau $ is small, one can linearise Eq. \ref{eq:mu} assuming that the non-linear phase shift is small, i.e. by substituting $ \mu_e $ in the argument of $ \Delta \phi(\tau,\mu_e) $ by $ \mu_0=\mu(\phi_0-\Delta\phi(\tau,\mu_0))$, so that
\begin{equation}
\mu_{e,lin}=\mu(\phi_0-\tau(1-\mu(\phi_0-\tau)\beta))).
\label{eq:mulin}
\end{equation}
This approximation works quite well up to $\beta\tau\lesssim0.1$ (see Fig. \ref{f:mu}).   

The precession phases corresponding to blobs propagating in the picture plane (i.e. $ \mu=0 $), given by  
\begin{equation}
\phi_{\perp,\pm}=\pm\arccos\left(\frac{-\cos\Theta_p\cos i}{\sin\Theta_p\sin i}\right)/2\pi\approx\pm0.34,
\end{equation}
are obviously not affected by light-travel-time effects at all and can
be considered stationary points for the considered
transformation. Interestingly, if the blobs get re-brightened at a
sufficiently long distance from the central source, it is possible
that photons emitted by an approaching ($ \mu>0 $) blob will reach the
observer before photons produced by a blob moving in the picture plane
($\mu=0$) despite the former having been launched later than the
latter. This can cause ambiguity in the expected position of the
observed X-ray line for time delays larger than a critical value that
can be found by minimising the expression
\begin{equation}
\tau=\frac{1}{\beta\mu}\left[\arccos\left(\frac{\mu-\cos\Theta_p\cos i}{\sin\Theta_p\sin i}\right)/2\pi-\phi_{\perp,+}\right]
\end{equation}
over $ \mu$ in the range from $ 0 $ to $\mu_{max}=\cos(i-\Theta_p)$. For the parameters of the SS 433 jets, we find $ \tau_1\approx1.8$ at $ \mu\approx0.28 $. 

The distance for which photons emitted by a blob launched at the phase
$ \phi=0 $ (and having the maximum $ \mu $) begin to catch up with
photons from a blob launched earlier at the phase $\phi_{\perp,-} $ is given by
\begin{equation}
\tau_2=\frac{1}{\beta\mu_{max}}\left[\arccos\left(\frac{\mu_{max}-\cos\Theta_p\cos i}{\sin\Theta_p\sin i}\right)/2\pi-\phi_{\perp,+}\right],
\end{equation}
which yields $ \tau_2\approx2.35$ for SS 433 parameters, and this
marks the instant when half of all approaching blobs are involved in
the ambiguity of the predicted line position. 

Further, $ \tau $ for which the arrival time of photons from the
$ \mu=\mu_{max} $ jet's regions coincides with the arrival time of
photons emitted by a blob launched at the phase $\phi_{\perp,+}$ of
the preceding precession cycle equals 
\begin{equation}
\tau_3=\frac{1}{\beta\mu_{max}}\left[\mu_{max}-\arccos\left(\frac{\cos\Theta_p\cos i}{\sin\Theta_p\sin i}\right)/2\pi-1+\phi_{\perp,+}\right],
\end{equation}
{after which more than half of all points of the initial precession
  curve are involved in the ambiguity of the predicted line position), and $
  \tau_3\approx4.4$ for SS 433. All these situations are illustrated
  in Fig. \ref{f:munl}. Finally, full 'phase mixing' takes place at $
  \tau\approx2\tau_3\sim9$. 

%
\begin{figure}
\centering
\includegraphics[width=0.4\textwidth,bb=80 200 530 680]{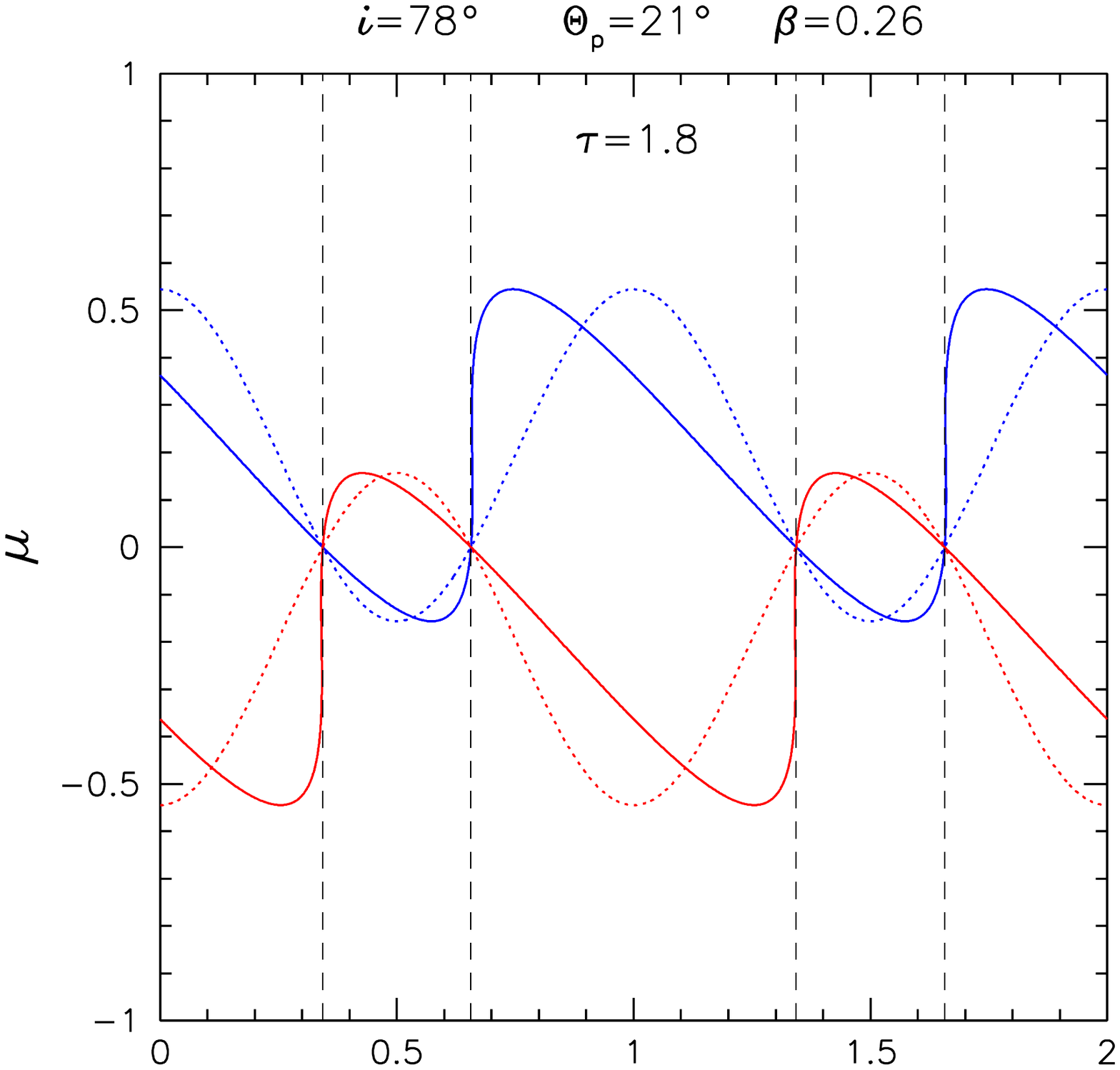}
\includegraphics[width=0.4\textwidth,bb=80 200 530 650]{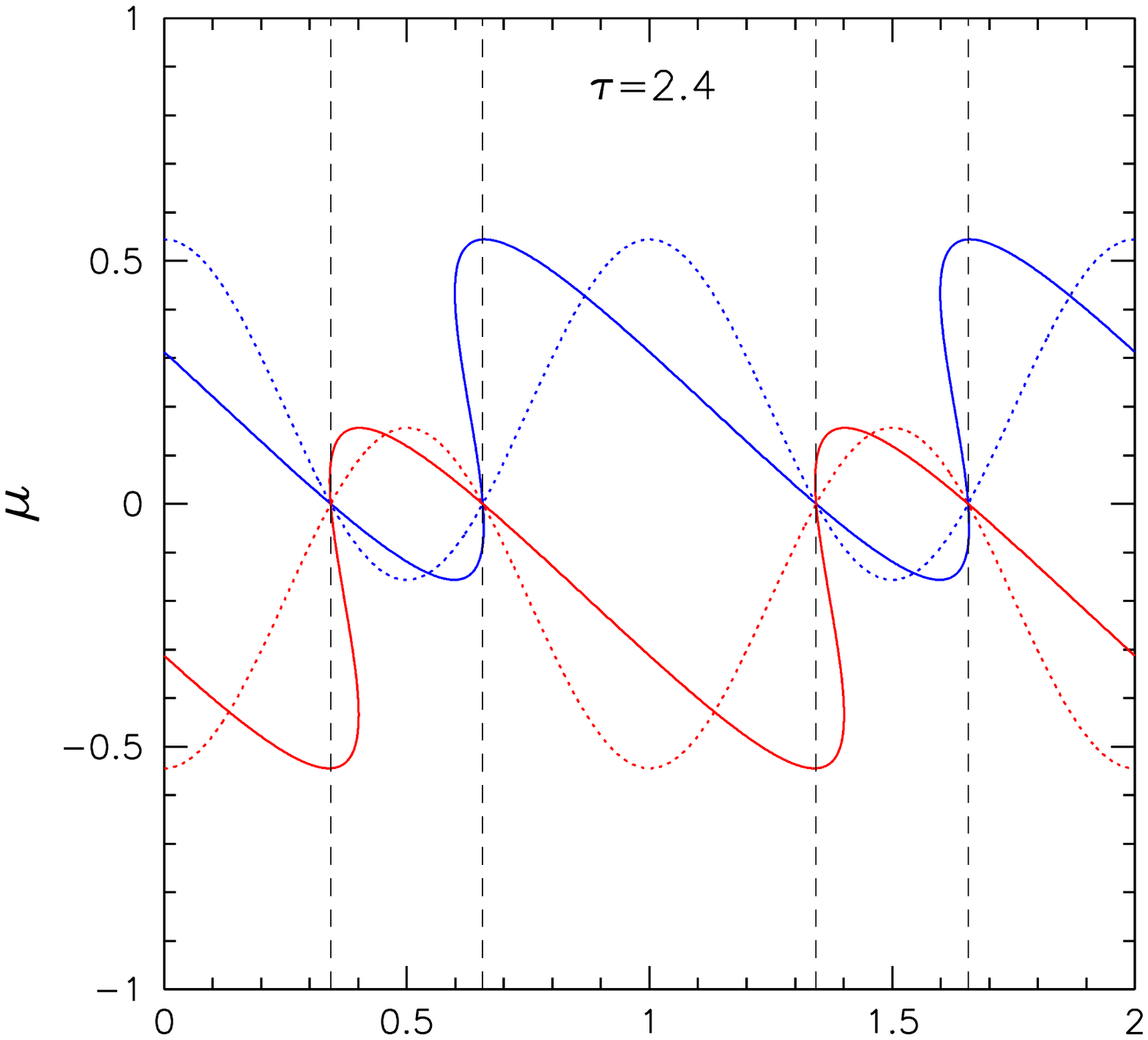}
\includegraphics[width=0.4\textwidth,bb=80 180 530 650]{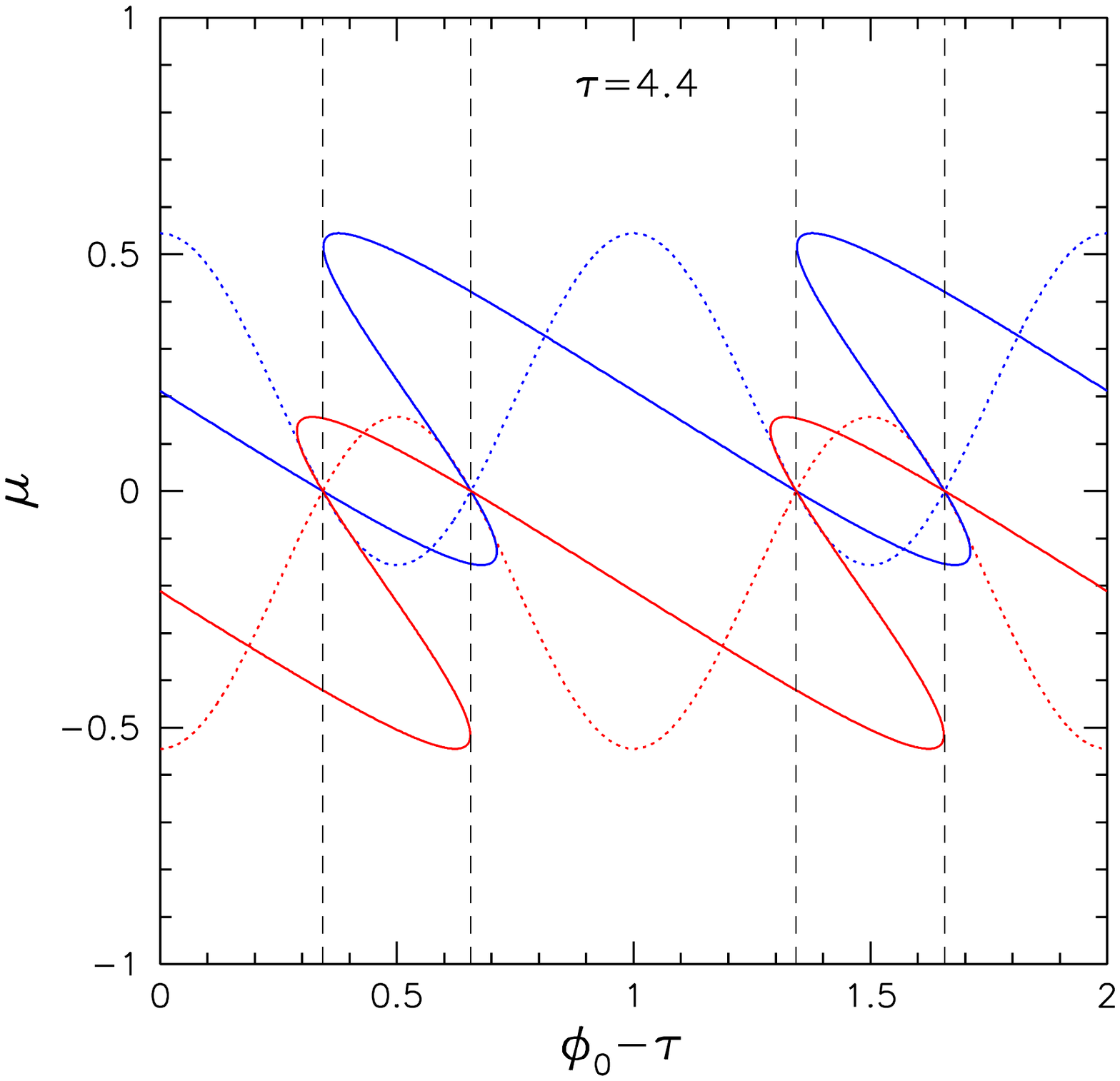}
\caption{\small The same as Fig. \ref{f:mu} but for the regime when photons from several blobs launched at different precession phases arrive at the observer at the same time, giving rise to ambiguity in the predicted X-ray line positions. The values of $ \tau$ are 1.8, 2.4 and 4.4 (from top to bottom).}
\label{f:munl}
\end{figure}
%

The steepening of the $\mu(\phi_0)$ curve at $ \phi_{\perp,-}\approx0.66 $ for  $\tau$ close $ \tau_1\approx1.8 $ (and at $ \phi_{\perp,+}\approx0.34$ for the red jet) means that significant brightening of the jet emission must occur at this phase (if the emission is spatially unresolved), since blobs that were launched over $\approx20\%$ of the precession period will become bright at this moment simultaneously (see Fig. \ref{f:munl}). The arising X-ray line will be significantly broadened. The same effect should also take place for other $ \tau $ but the associated 'amplification factor', 
\begin{equation}
A(\tau,\phi_0)=|\left(\frac{d\phi_e(\tau,\phi_0)}{d\phi_0}\right)^{-1}|=|\left(1-\frac{d\Delta\phi(\tau,\phi_0)}{d\phi_0}\right)^{-1}|,
\end{equation}
is smaller (see Fig. \ref{f:ampl}).

In the regime when a set or range of $\phi_e$ values, $ [\phi_e] $, correspond to the same $\phi_0$, one needs to perform summation (integration) over this set (range) with the weights of different $ \phi_e $ determined by the square of the Doppler boosting factor $ \delta(\mu_e)$ (for the photon intensity from a blob with given $ \mu_e $).

 Thus, some property $S$ of the observed X-ray line can be calculated as 
\begin{equation}
\tilde{S}(\tau,\phi_0)=\int_{[\phi_e]}d\phi_e~A_{\phi_e}(\tau,\phi_0)~\delta^2(\mu_e(\phi_e))~S(\phi_e)/w_0,
\end{equation}
where
\begin{equation}
w_0(\tau,\phi_0)=\int_{[\phi_e]}d\phi_e~A_{\phi_e}(\tau,\phi_0)~\delta^2(\mu_e(\phi_e))
\end{equation}
is the normalising factor.

%
\begin{figure}
\centering
\includegraphics[width=0.4\textwidth,bb=70 200 530 680]{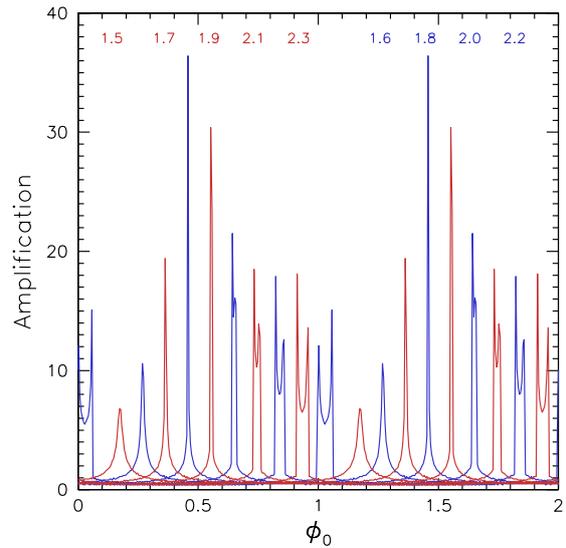}
\caption{\small Amplification factor for the intensity of the
  unresolved emission due to simultaneous arrival of photons emitted
  by parts of the jet launched at different precession phases, as a
  function of the precession phase of the central source, for
  different values of the time-delay parameter $ \tau $ as indicated
  by the numbers above each curve's peak (coloured alternately blue
  and red for better visibility). A characteristic double-peaked
  structure emerges for $ \tau>1.8$ reflecting two 'turnover' points
  in the line position curve (see Fig. \ref{f:munl}). } 
\label{f:ampl}
\end{figure}
%

Let us now consider a more general situation when the jet's matter
becomes X-ray bright not at one particular $ \tau $, but for $ \tau $
ranging from $ \tau_s$ to $ \tau_f$, and one cannot spatially resolve
the signal corresponding to different $\tau$ and $ \phi_e $ and
measures some integral characteristics of the line $<S> $ (primarily,
position and the total intensity) in the spectrum of such emission,
which can be calculated as
\begin{equation}
<S>_{ \tau_s,\tau_f}(\phi_0)=\int^{\tau_f}_{\tau_s}d\tau w_0(\tau,\phi_0)\tilde{S}(\tau,\phi_0)/W_0, 
\label{eq:long}
\end{equation}
with the normalising coefficient
\begin{equation}
W_0=\int^{\tau_f}_{\tau_s}d\tau w_0(\tau,\phi_0).
\end{equation}
Figure \ref{f:mut} shows the resulting effect for various $ [\tau_s,\tau_f]$ intervals defined by their width $ \Delta\tau=\tau_f-\tau_s $ and the central point $<\tau>=(\tau_s+\tau_f)/2$.  

As could be expected, the amplitude of the observed modulation with
precession phase decreases with increasing width of the averaging
window $ \Delta\tau $. However, the modulation remains noticeable even
for $\Delta\tau=0.9$, due to the Doppler boosting on the segment of the
modulation curve with the steep variation in the predicted line
centroid energy. For the receding jet, the situation is inverted,
which results in a significantly smaller amplitude of the modulation
of the line centroid with precession phase. 

Since $\tilde{S}(\tau,\phi_0)$ is given by integration over $ \phi_e
$, one can change the integration order in Eq. (\ref{eq:long}), namely
first average the amplification factor  $ A_{\phi_e}(\tau,\phi_0) $
over $\tau$ and thus significantly weaken its influence on the net
result.  For the duration of the brightening period $ \Delta\tau >
(\mu_{max}-\mu_{min})\beta\tau\approx 0.2\tau $, the effect of the
relative amplification is already quite small, so the exact solution
is well approximated by the prediction that only includes Doppler
boosting when integrating over the whole emitting region (see the
dash-dotted line in Fig. \ref{f:mut}).   

Before comparing our predictions with the observed line positions, it
is worth mentioning that since $ \mu\beta  <<1 $, $ \delta(\mu) $ is
well approximated by the linear function $(1+\beta\mu)/\gamma $, which
implies that the predicted variation of the line position $
E=\delta(\mu_e)E_{0}$ must closely follow the dependence of $ \mu_e $
on the precession phase. 

\begin{figure}
\centering
\includegraphics[width=0.4\textwidth,bb=80 200 530 680]{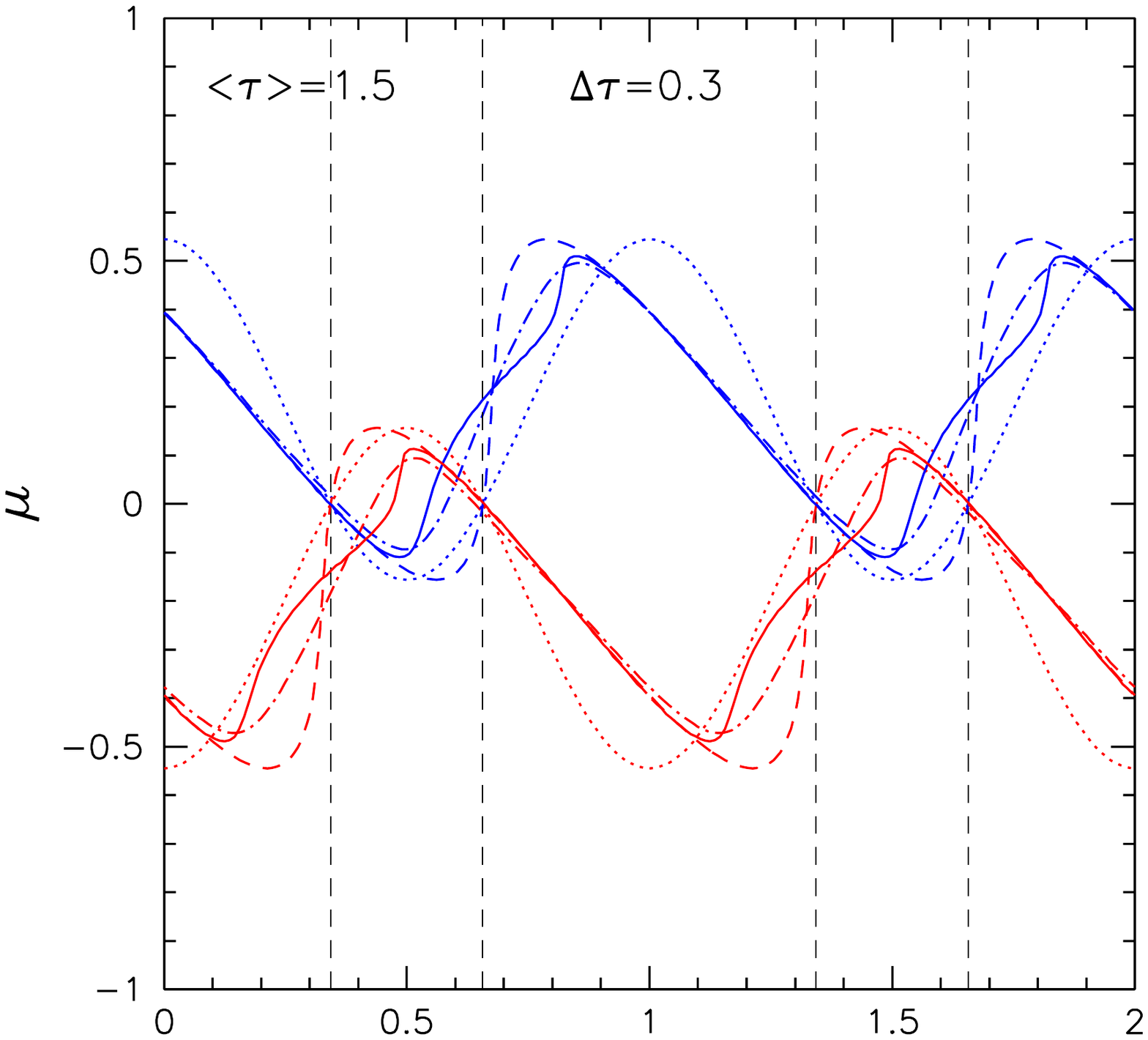}
\includegraphics[width=0.4\textwidth,bb=80 200 530 650]{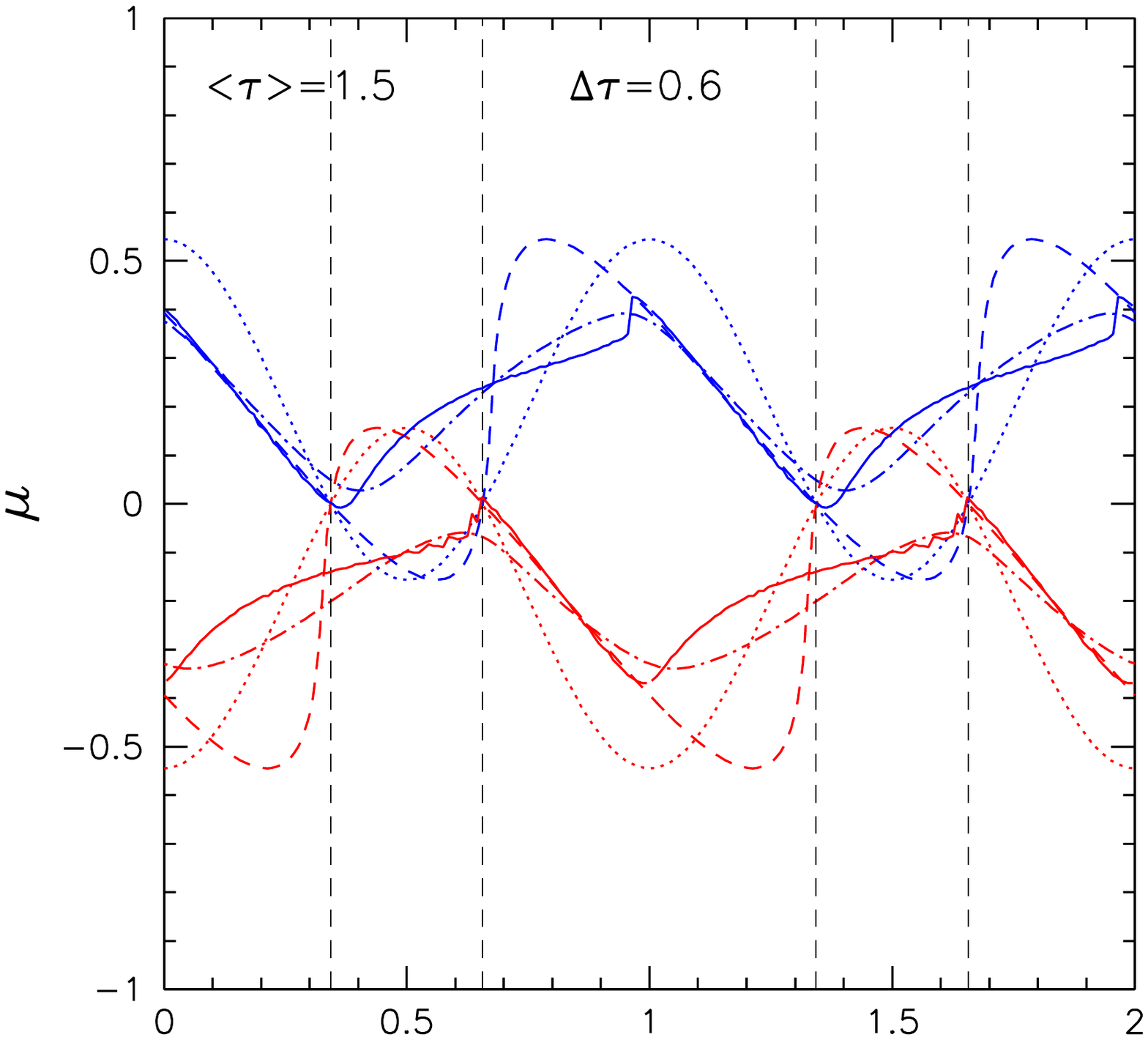}
\includegraphics[width=0.4\textwidth,bb=80 180 530 650]{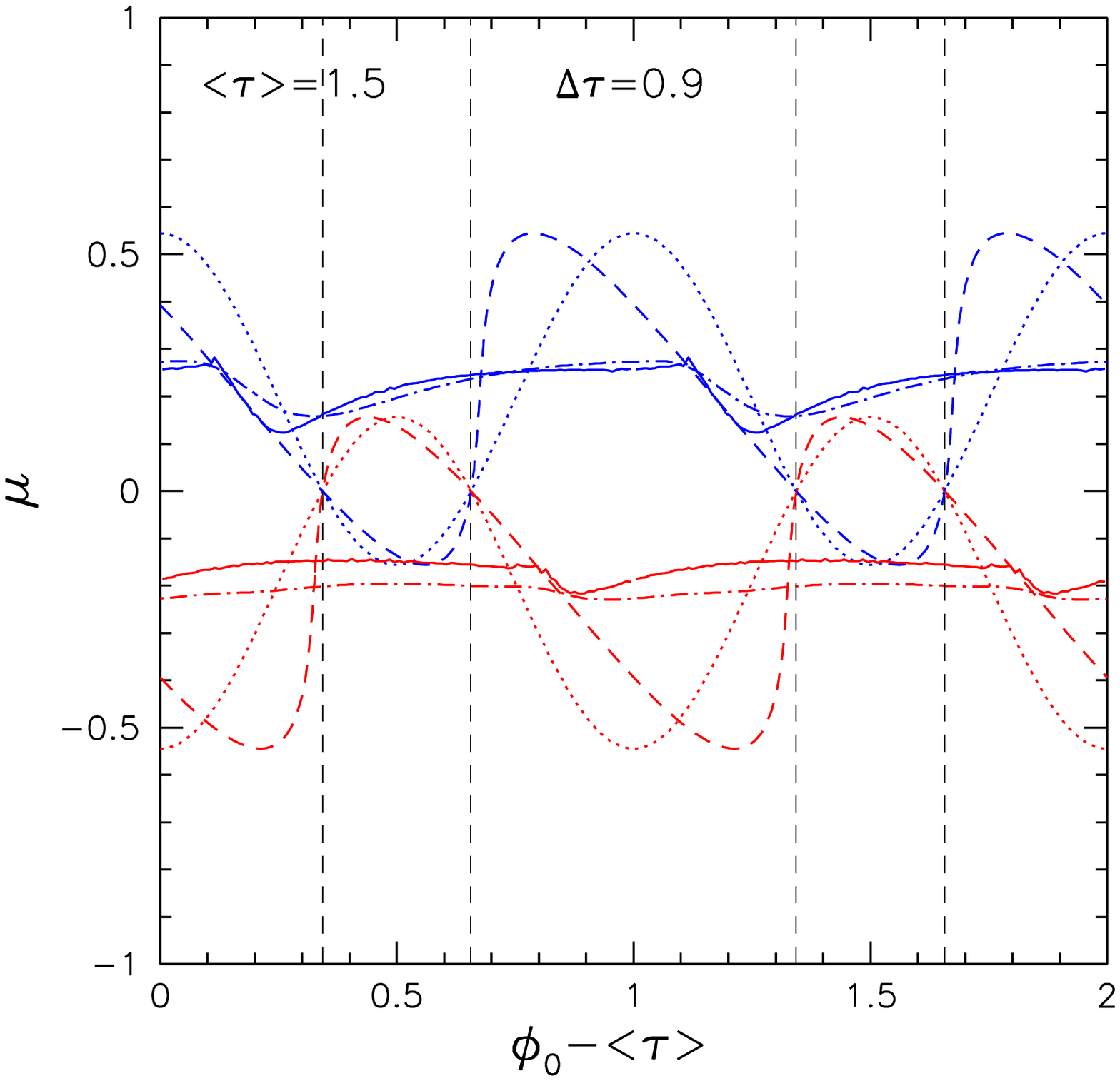}
\caption{\small Averaged cosine of the angle between the jet's
  direction and the line of sight for the scenario of long-duration
  brightening as a function of the central source's precession phase
  shifted by $<\tau>$. The blue and red solid lines show the
  predictions for the blue and red jets respectively, taking into
  account both Doppler boosting and arrival time overlaps, while the
  dash-dotted lines show the prediction accounting for Doppler
  boosting only. The predicted line positions for $\tau=<\tau>$ in the
  short flare scenario with and without allowance for the
  light-travel-time effects are shown with dashed and dotted lines,
  respectively.
}
\label{f:mut}
\end{figure}
\section{Comparison with the data}
\label{s:comparison}
%

\cite{Migliari2002} reported the detection of X-ray emission lines at
energies $ E_{b,o}=7.28^{+0.02}_{-0.23} $ keV and $
E_{r,o}=6.39^{+0.12}_{-0.15} $ keV, apparently associated with the
blue and red jets, respectively, at distances $ \approx
1.5\times10^{17}$ cm $\approx1.4\lambda$ (in projection) from the
central source (see also \cite{Migliari2005}), which translates to $
\tau =1.4-1.6$ depending on $ \mu_e $. Since the duration of the
relevant (as well as subsequent ones) \textit{Chandra} observation
of the arcsec-scale X-ray emission is a few tens of kiloseconds
(see Table 1 in \cite{Migliari2005}), i.e. less than $ \sim0.1 $\% of
the precession period, these observations can be considered
instantaneous snaphots of the extended jets. The SS 433 precession
phase for the \cite{Migliari2002} observation was $\phi_0=0.63 $ (see
Table 1 in \cite{Migliari2005}, but note the 0.5 shift between the
definition of the zero phase in their work and the one adopted here). 

Below, we first check whether the observed X-ray line positions can be reconciled with the predictions of the kinematic model taking into account light-travel-time effects. We consider four possible line identifications: Fe XXV K$\alpha$ (6.7 keV) and Fe XXVI Ly$\alpha$ (6.96 keV) -- the two brightest iron lines from hot optically thin plasma; and Fe I K$\alpha$ (6.4 keV) and Fe I K$\beta$ (7.06 keV) -- the brightest fluorescent lines of neutral iron.

\begin{figure}
\centering
\includegraphics[width=0.35\textwidth,bb=50 200 550 670]{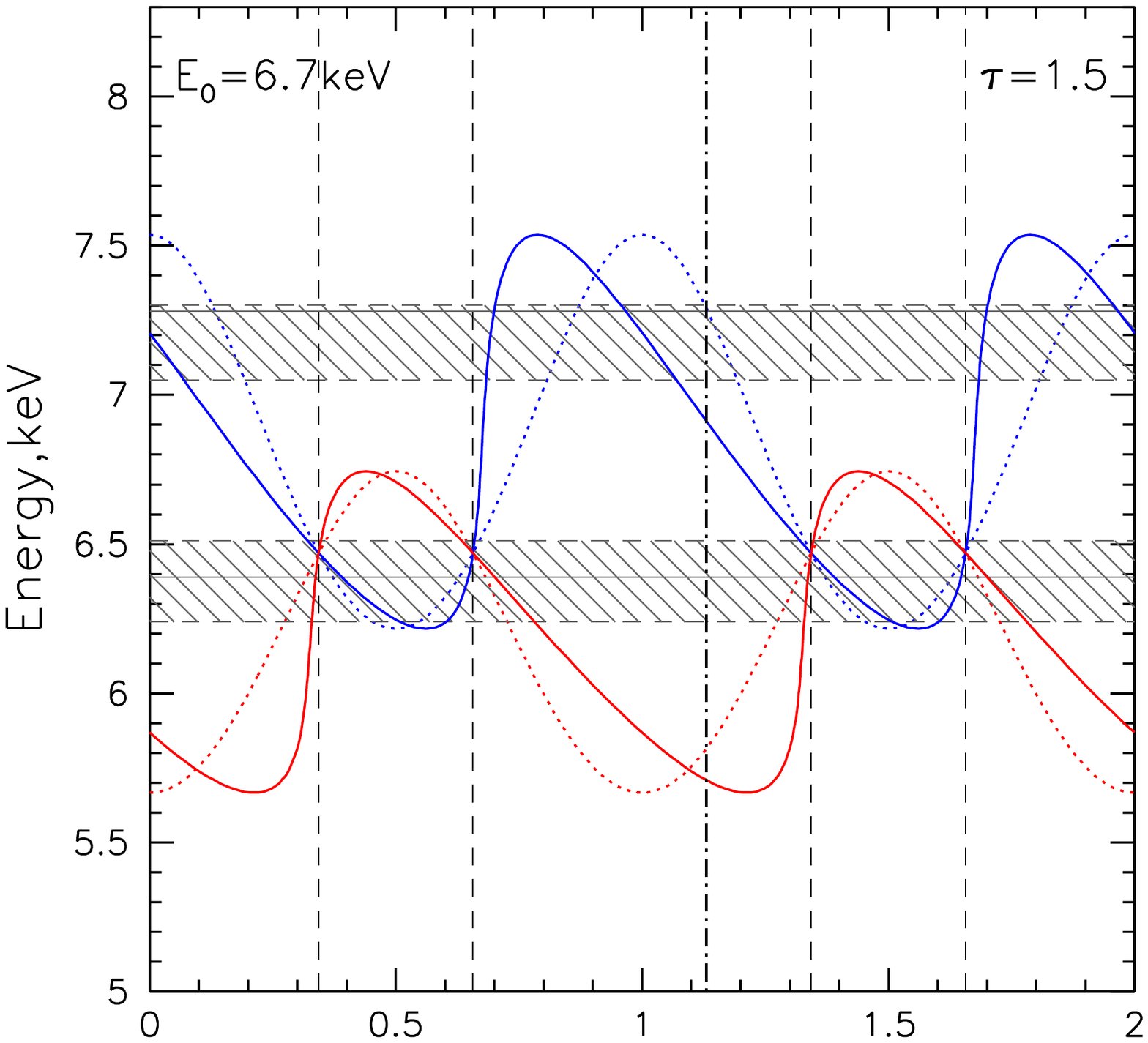}
\includegraphics[width=0.35\textwidth,bb=50 200 550 650]{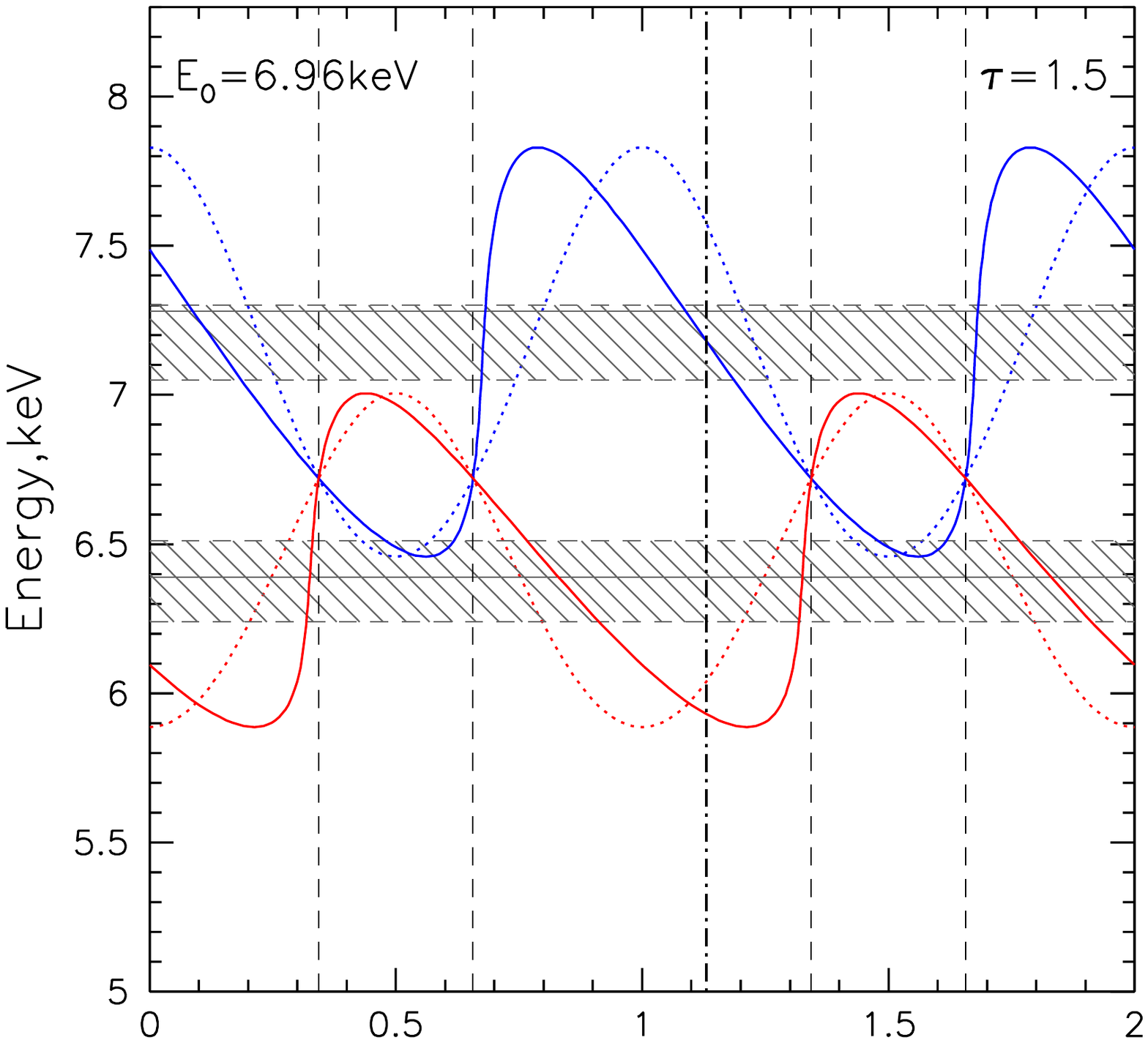}
\includegraphics[width=0.35\textwidth,bb=50 200 550 650]{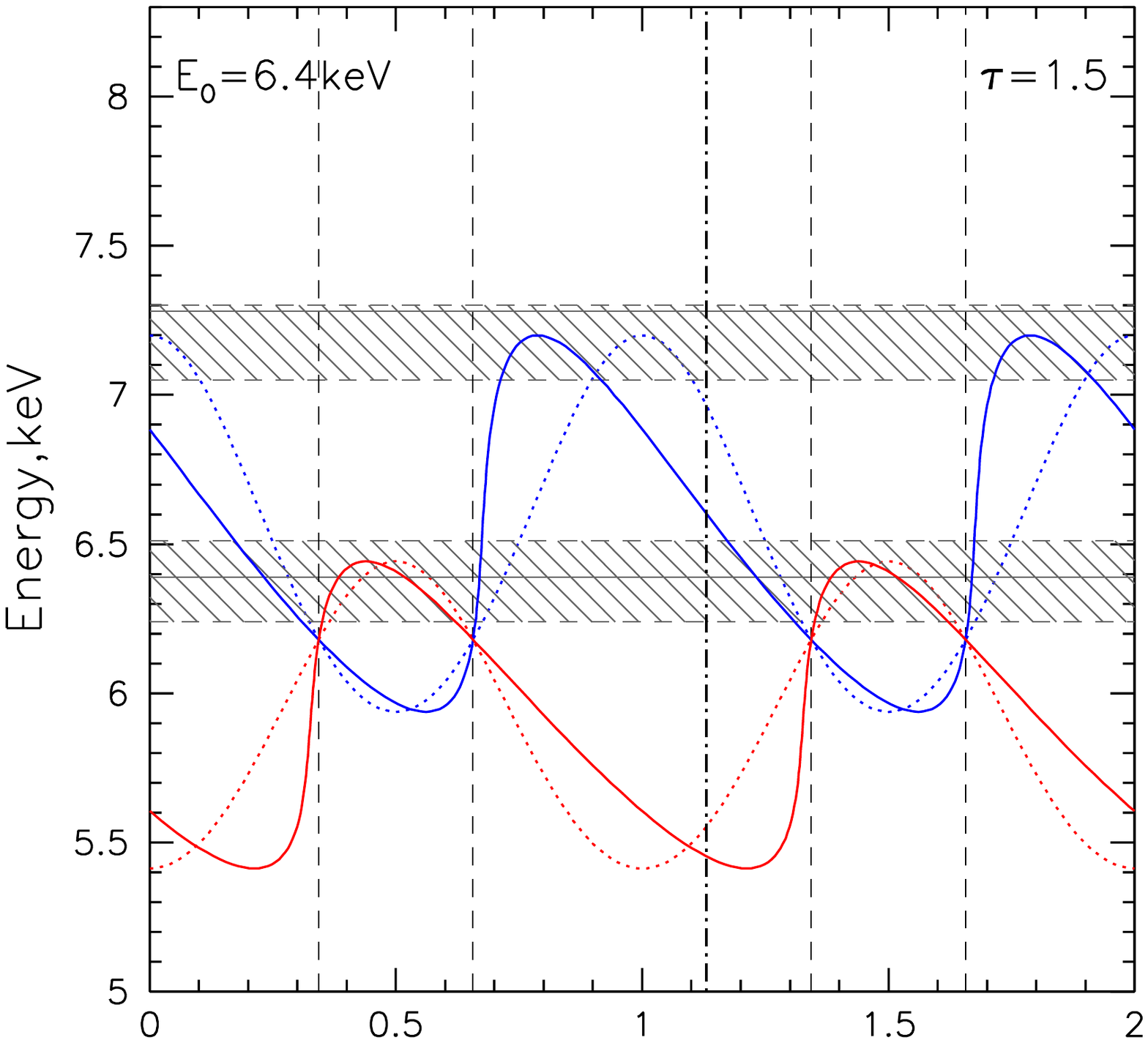}
\includegraphics[width=0.35\textwidth,bb=50 180 550 650]{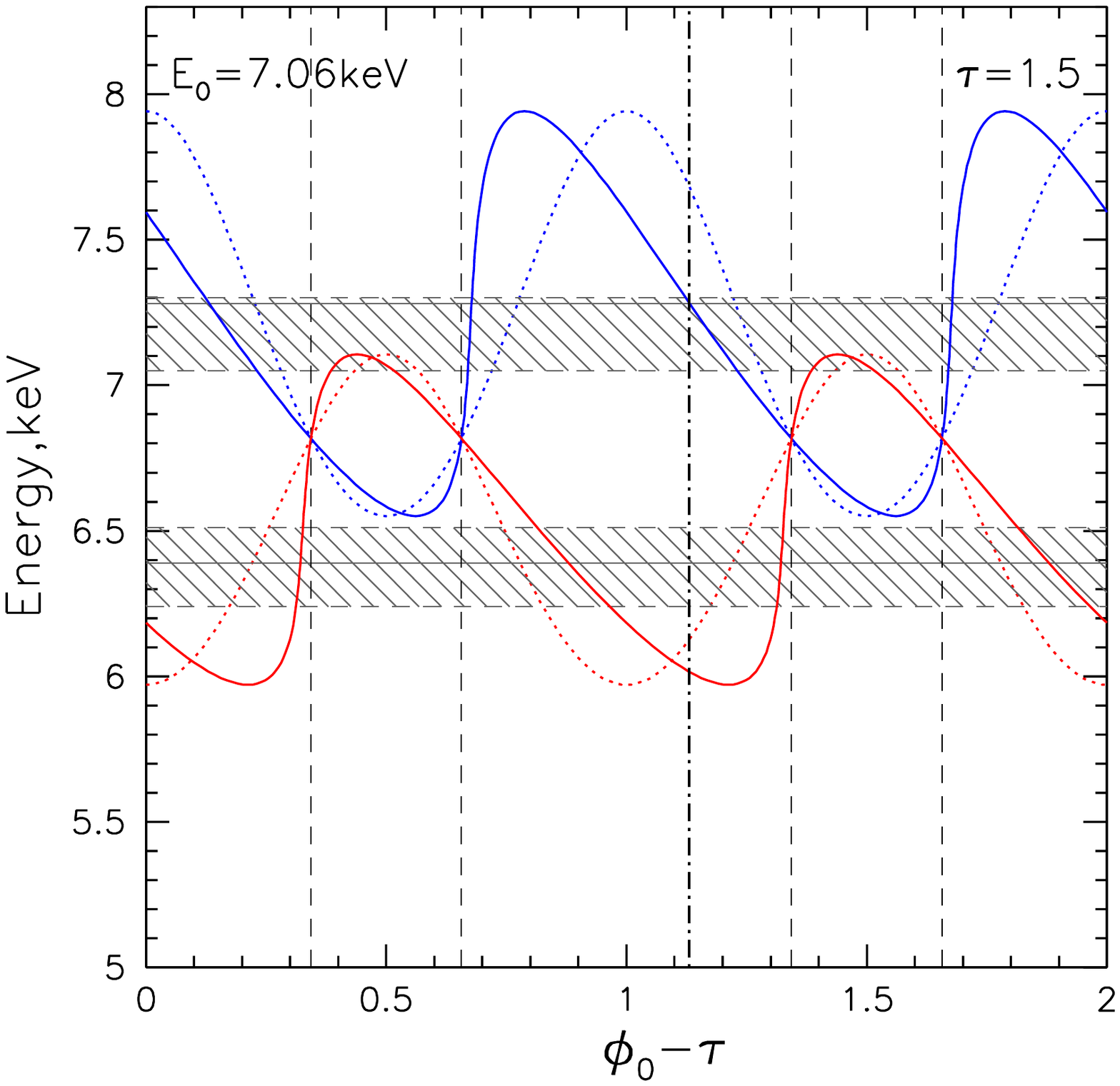}
\caption{\small Predicted line positions (solid lines) for the blue
  and red jets in the short brightening scenario with $ \tau=1.5$. The
  line's rest-frame energy $ E_0 $ equals to 6.7, 6.96, 6.4 and 7.06
  keV (from top to bottom). The dotted line shows the prediction of
  the kinematic model without allowance for light-travel-time
  effects. The hatched regions depict the 1$ \sigma$ uncertainty
  regions for the line centroids measured by \cite{Migliari2002}
  (solid horizontal lines) at the precession phase marked by the
  vertical dash-dotted line.} 
\label{f:ee}
\end{figure}

Figure~\ref{f:ee} shows the predicted variations of the positions of
these four lines in the short brightening scenario. Evidently, for $
\tau\approx1.5$, the observed X-ray line positions are consistent only
with the Fe XXV K$\alpha$ line at a phase close to $\phi_0-\tau
\approx\phi_{\perp,-}$, i.e. $ \phi_0\approx0.2 $. For Fe I K$\alpha$,
agreement is possible only for $ \tau>\tau_2=2.4$, which, however,
contradicts the observed location of the X-ray emission at $ \tau<2
$. Similarly, for the Fe XXVI Ly$\alpha$ and Fe I K$\beta$ lines, a
match is possible for $ \tau\approx3.5 $ or $ \tau\approx0.2$, thus also
inconsistent with the observations. 

In order to quantify the quality of the model fit to the observed line
positions, let us introduce the statistics 
\begin{equation}
\chi^2(\tau,E_0)=\left(\frac{E_{b,m}-E_{b,o}}{\sigma(E_{b,o})}\right)^2+\left(\frac{E_{r,m}-E_{r,o}}{\sigma(E_{r,o})}\right)^2,
\label{eq:chi}
\end{equation}
where $E_{b,m}$ and $E_{r,m}$ are the model predictions for the blue
and red jets, and  $ \sigma(E_{b,o}) $ and $ \sigma(E_{r,o}) $ are the
upper and lower uncertainties for the measured values.

Projection of the parameter space region encompassed by the $
\Delta\chi^2(\tau,E_0)=1$ contour on the $E_0$ axis with $ \tau $ in
the range from 1.1 to 1.8 gives $ E_0\approx 6.7\pm0.2$ keV, where
the uncertainty is $ 1\sigma$. Thus, in the short brightening
scenario, the observed line positions are consistent with the
prediction for the Fe XXV K$\alpha$ line, whereas the other considered
lines cannot provide fits of comparable quality. However, for $
E_0=6.7 $ keV, the best fitting phase $ \phi_0\approx 0.2 $ is in 
disagreement with the precession phase of the central source for the
\textit{Chandra} observation. This motivates us to consider the more
complicated long brightening scenario.
    
The predicted line positions in the long brightening scenario are
shown in Fig. \ref{f:eet} for $ <\tau>=1.5$ (i.e. $r\approx 1.6\times
10^{17}$ cm from the central source) and $ \Delta\tau=0.3,0.6,$ and $
0.9$ (corresponding to the emission region's size of $\approx
3.3\times 10^{16}$, $6.55\times 10^{16}$ and $9.8\times 10^{16}$ cm,
respectively). We see that for $ <\tau>=1.5$, almost no dependence on
$\Delta\tau$ is expected for the predicted line position for the blue
jet at the precession phase corresponding to the observation
discussed. This allows us to exclude both the $ E_0=6.4$ keV and $
E_0=6.7$ keV lines as possible identification candidates in this
scenario. Our predictions for the $ E_0=6.96$ keV and $ E_0=7.06$ keV
lines are broadly consistent with the observations for $\Delta\tau$
from $ \approx $0.45 to $\approx$1 in the former case and $\Delta\tau$
from $ \approx $0.4 to $\approx$0.75 in the latter. The best fitting
$\Delta\tau$ is $\approx0.6$ in both cases. 

The predicted line widths calculated as the dispersion of the
predicted line positions for emission coming from different blobs
observed simultaneously equals $ \sigma(E)\approx 0.3$ keV for both
the blue and red jets if $\Delta\tau=0.6$. For $\Delta\tau=0.9$, $
\sigma(E)\approx 0.5$ keV and $ \sigma(E)\approx 0.4$ keV for the blue
and red jets, respectively. Our results are thus broadly consistent
with the results of the simulation performed by \cite{Migliari2002},
who found $ E_0=7.06$ keV for the $\Delta\tau=1$ scenario, with the
predicted line widths of $ \sim $ few $ 0.1 $ keV.

\begin{figure}
\centering
\includegraphics[width=0.35\textwidth,bb=50 200 550 670]{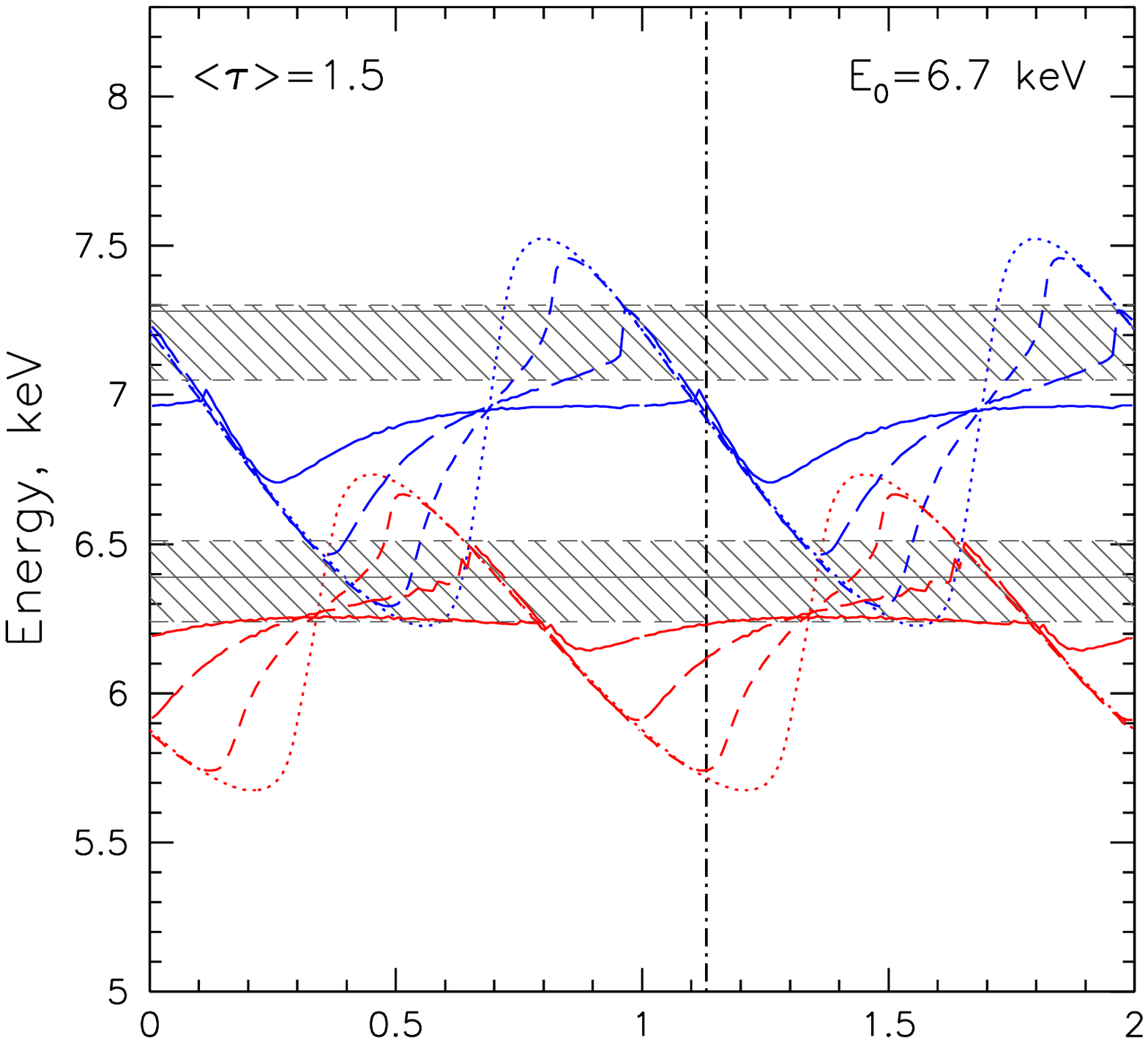}
\includegraphics[width=0.35\textwidth,bb=50 200 550 650]{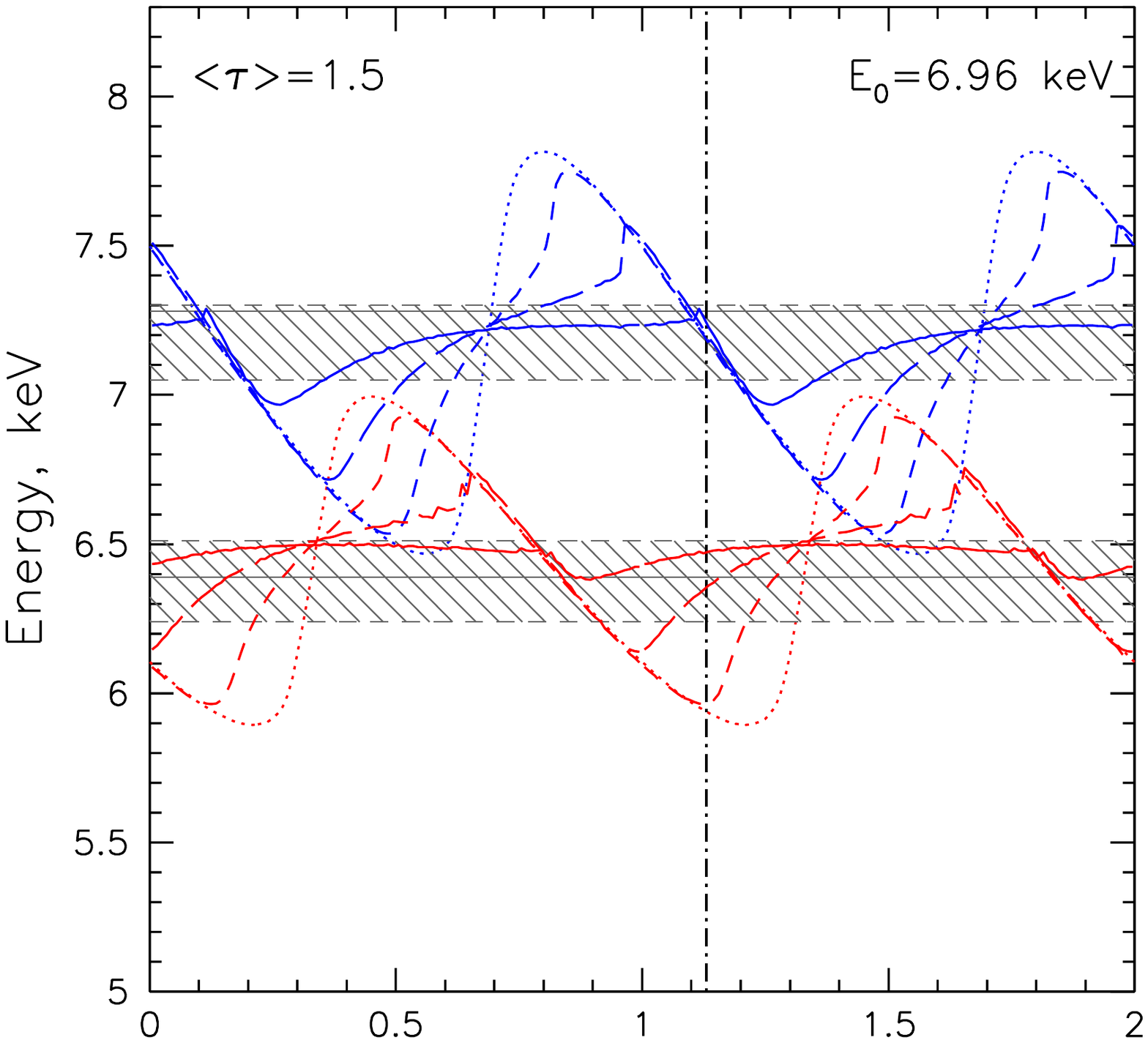}
\includegraphics[width=0.35\textwidth,bb=50 200 550 650]{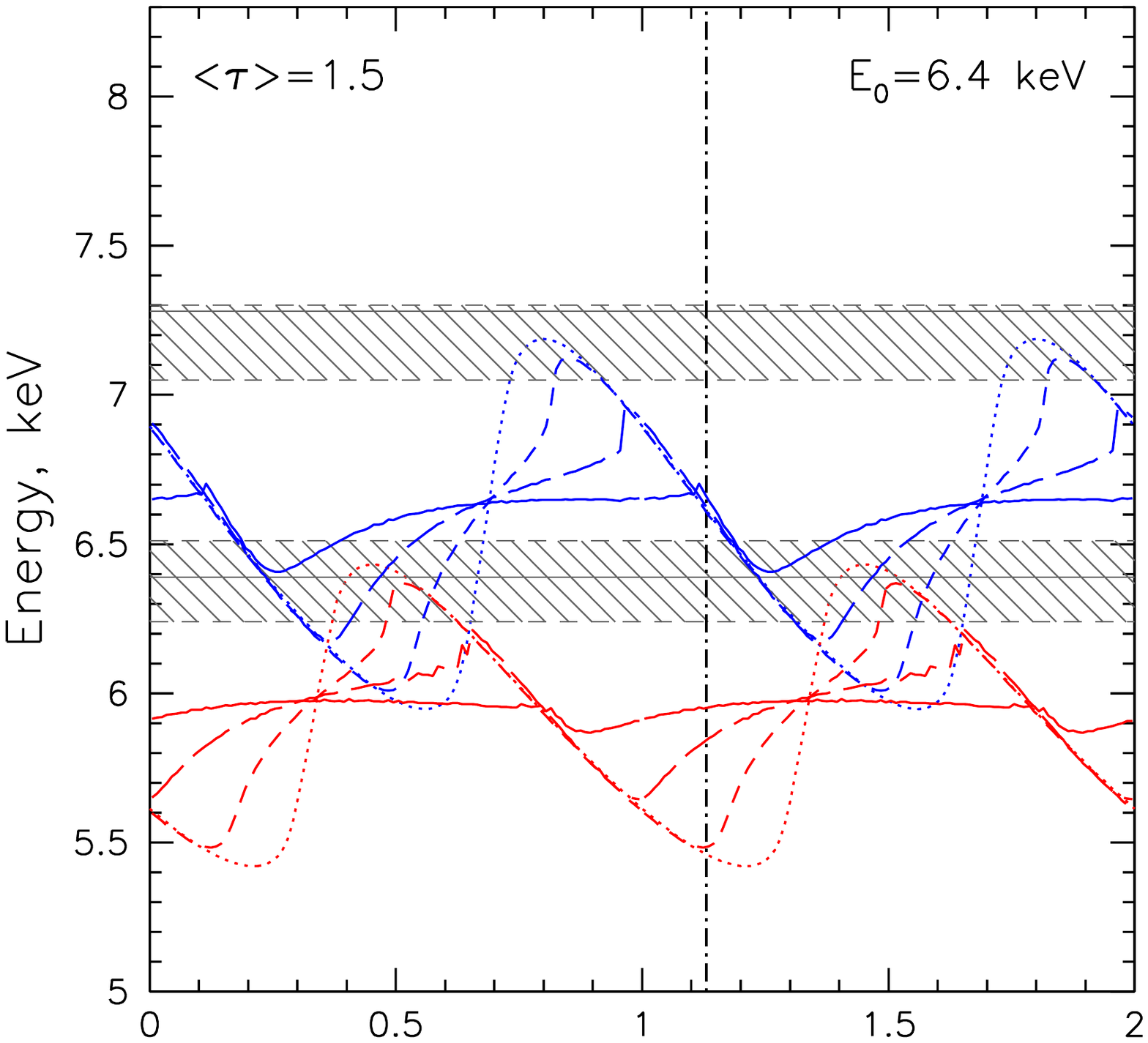}
\includegraphics[width=0.35\textwidth,bb=50 180 550 650]{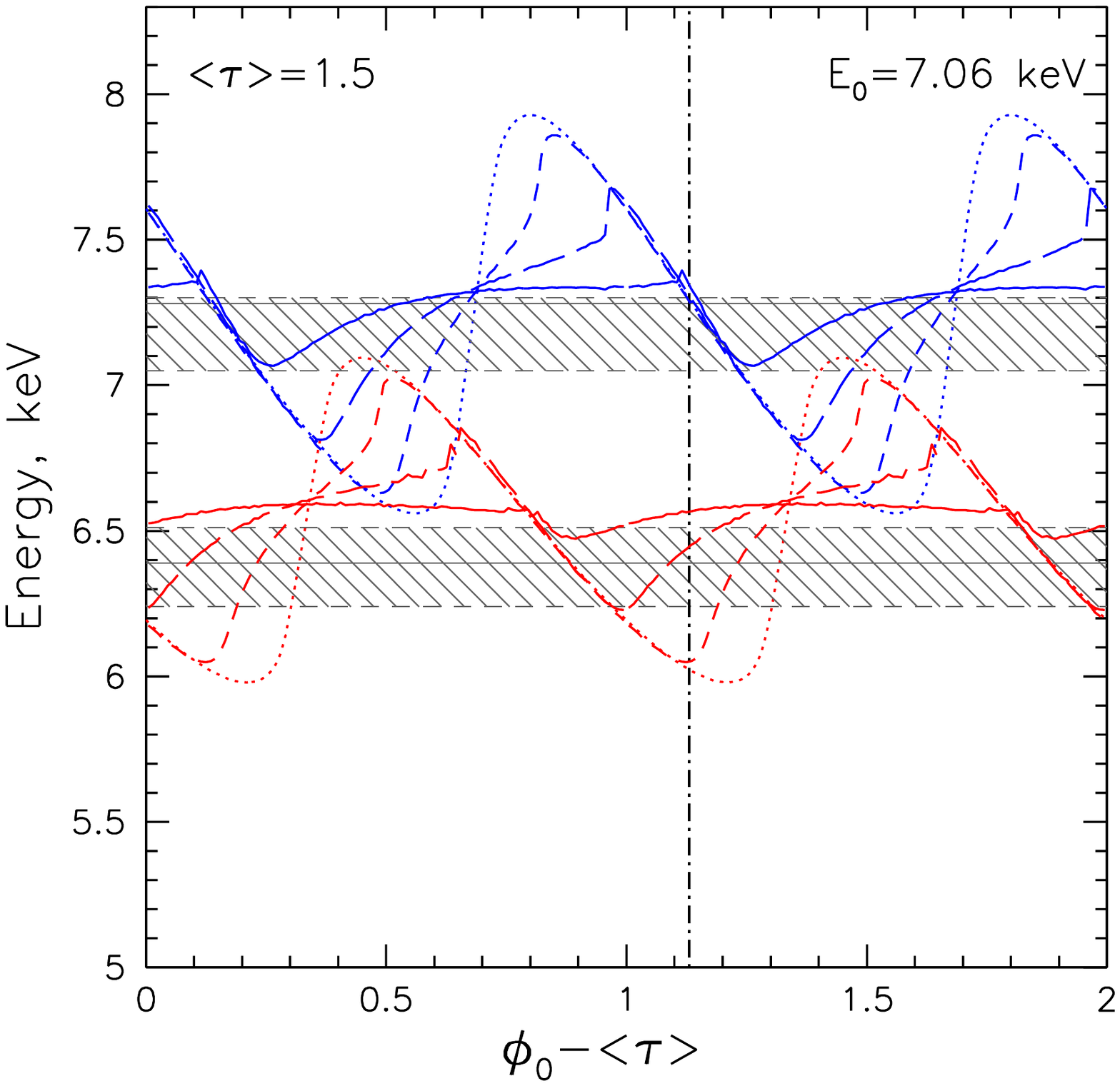}
\caption{\small Same as Fig.\ref{f:ee} but for the scenario with a long brightening period centered at  <$\tau$>=1.5  and having duration $d\tau=\tau_f-\tau_f$ equal to 0.3 (short-dashed line), $0.6$ (long-dashed line), and $0.9$ (solid line). For comparison, the prediction for the short flare scenario with $ \tau=<\tau>$ is shown with the dotted line.}
\label{f:eet}
\end{figure}

\section{Discussion and conclusions}
\label{s:discussionconslusions}
%

\cite{Migliari2002} proposed that the observed X-ray lines are
associated with the 7.06 keV, i.e. Fe I K$\beta$, line. However, such
a line can arise as a result of fluorescence of cold gas irradiated by
X-rays, rather than from a hot ionized plasma as was suggested by
\cite{Migliari2002}. Moreover, if the observed lines were indeed
associated with Fe I K$\beta$, there must also be $ \sim 5-10$ times
stronger Fe I K$\alpha$ lines present in the spectrum, which are not
observed. We have demonstrated that the observed line positions are in
fact consistent with the Fe XXVI Ly$\alpha$ (6.96 keV) line, with 
different Doppler shifts for the blue and red jets. Physically,
this implies that the emitting plasma should be re-heated up to
temperatures $ \gtrsim 10$ keV at $\sim 10^{17}$~cm from the central
source in a region of size $\sim 6\times 10^{16}$ cm. If the
temperature of the emitting plasma were somewhat lower, the centroids
of the observed lines would be shifted to lower energies as a result
of the contribution from the Fe XXV K$\alpha$ (6.7 keV) line to the
spectrally unresolved broad emission line.    

\cite{Migliari2002} proposed that internal shocks arising due to variation in the speeds of blobs launched in approximately the same direction, i.e. with a time delay of $ \approx P$, can provide a mechanism for the jets' re-heating. Based on the detected short-term (on the time-scale of hours and days) variability, \cite{Migliari2005} further hypothesised the existence of an underlying outflow of highly energetic particles driving a fast shock wave. Another mechanism can be related to heating of the jet's matter as a result of interaction with the surrounding cocoon, which causes an oblique shock propagation inside the jets \citep{Heavens1990}.

At distances of $ \sim 10^{17} $ cm from the central source, the
matter of the jets could in fact be composed of a large number of
dense blobs, similarly to what is believed to occur in the region of
SS 433 optical line emission \citep{Fabrika2004}. From the energetic
considerations, the number density of the gas responsible for the
extended X-ray emission needs to be quite high: 
\begin{equation}
n\sim 1\times 10^6~L_{X,33.5}/N_{50}/\Lambda_{-22.5} ~ {\rm cm^{-3}},
\end{equation}
where $ L_{X,33.5}\approx1$ is the observed luminosity of the extended
X-ray emission in units of $ 10^{33.5} $ erg/s \citep{Migliari2002}, 
$\Lambda_{-22.5}\approx1$ is the plasma emissivity in the same spectral
band for $ T\sim10  $ keV in units of $ 10^{-22.5} $ erg/s cm$^3$, and
$ N_{50}=k\dot{M}_j\Delta\tau P/m_p/10^{50}\sim k\Delta\tau
$ is the total number of hydrogen nuclei (divided by $10^{50}$) in the emitting region of
size $ \Delta\tau $ (along the jet) for the mass flux through the jet 
$ \dot{M}_j\sim10^{-6.5}~M_{\odot}/$yr \citep{Fabrika2004}, and $k$ is
the mass fraction of the emitting gas.  

The mean density of a continuous jet at this distance would be $ \sim
10^{2} $ cm$^{-3}$ (assuming the X-ray jets in the compact core are
characterized by the size $\sim 10^{11}$ cm and number density $\sim
10^{14}$ cm$^{-3}$, e.g. \citealt{Marshall2002, KMS2016}), i.e. less than the
required density by at least 4 orders of magnitude. If the length of
the X-ray bright period $ \Delta\tau P$ is determined by the cooling
time of the emitting matter $\sim T/(n\Lambda)$, then it is easy to
find that the required $ k$ is $\approx 0.03$, i.e. 3 \% of the total
mass of the jet's matter in this region. This mass could be provided
by a number of dense blobs experiencing collisions as a result of more
efficient deceleration of leading blobs by the interstellar medium
compared to trailing ones in a regime where the boundary of the jet
becomes virtually perpendicular to the local gas velocity direction,
i.e. at distances comparable with $\lambda/\pi\tan\Theta_p\approx
0.8\lambda$ \citep{Heavens1990}. 

The characteristic collision velocities are $\sim\delta V\sim \Theta
_{j}\beta~c\sim 1000$ km/s, which should result in heating of the blobs
up to temperatures $ \sim 10 $ keV, similar to the observed
temperatures at the base of the X-ray jets near the central source
\citep{Marshall2002}. Blob collisions are possible only during the 
period $ \delta t_{rc}\sim r\Theta_{j}/\delta V\approx P $  
until a complete lateral re-collimation of the precessing jet takes
place and its shape becomes that of a band. After that, the X-ray
emission should gradually cease at the cooling time-scale. 

Based on the constraints on the origin of the observed lines obtained
here, namely the requirements of high temperature of the emitting
plasma and relatively long duration of the brightening episode $
\Delta \tau \sim 0.6$, and taking into account the observed rapid
(down to scales of several hours, \citealt{Migliari2005}) stochastic
variability of the X-ray emission, we suggest that dissipation of the
jets' kinetic energy via collisions of their discrete dense blobs due
to their deceleration by the medium of the jets' cocoons, provides a
natural explanation of the observed characteristics of the
arcsec-scale X-ray emission.

\section{acknowledgements}
The research was supported by the Russian Science Foundation (grant 14-12-01315).

\end{document}